\documentstyle[prb,aps,multicol,epsf]{revtex}

\ifpreprintsty\relax\else
\newlength{\colwidth}
\fi

\begin{document}
\draft 
\title{Distribution of Transmitted Charge through an Ultrasmall
  Double-Tunnel Junction}
\author{Heinz--Olaf M{\"u}ller\cite{email}} 
\address{Institut f{\"u}r Festk{\"o}rperphysik,
  Friedrich--Schiller--Universit{\"a}t, 
  Lessingstra{\ss}e 8, DE--07743 Jena, Germany} 
\date{\today}

\maketitle
\begin{abstract}
The transmission of charge through an ultrasmall double junction is
considered with Coulomb effects but at zero temperature. We construct an
equation which describes the time development of the transmission
probability of charges and solve this equation in terms of a recursion
relation. The results are compared with the double junction without
charging effects.
\end{abstract}

\ifpreprintsty\relax\else\begin{multicols}{2}\narrowtext\fi

\section{Introduction}

During more than one decade\cite{ave1,ful2} single-electron tunneling
(SET) has experienced considerable progress in both experiment and
theory. Basically, SET is a stochastic tunneling process of charges
occurring at ultrasmall low-capacitance high-resistance tunnel
junctions.\cite{ave1,ave3} Those tunnel junctions ensure that, 
on the one hand, the charging energy is large (compared to the
thermal energy) whereas, on the other hand, charge fluctuations are
suppressed. Thus, the number of tunneled charges forms a classical
variable whereas the tunneling probabilities (rates) are calculated
quantum mechanically, for instance in terms of Fermi's Golden
Rule.\cite{amm2} 

This calculation restricts the application of the semi-classical
approach to slow time dependencies on the scale of the $RC$--time of
the tunnel junction. It is in the range of $10^{-10}$s for nowadays
devices. There are approaches beyond the semi-classical theory
that allow the exploration of short time scales (for instance
Ref.~\onlinecite{sch8}), however for the purpose of this work simpler
Golden Rule approach is sufficient.

This paper is devoted to the study of the ultrasmall double tunnel
junction, which is one of the cornerstones of single electronics in
experiment and theory. It consists of two junction connected in series,
thus forming a small island inbetween. This system is capable
of displaying pronounced effects of single-electron tunneling like
Coulomb blockade, Coulomb staircase, and shot-noise suppression for
suitable parameters of the applied bias voltage $V$.

Throughout this paper zero temperature is assumed. On the one hand,
this assumption simplifies our consideration considerably and enhances
its clarity. On the other hand, many SET experiments are done at
sub-Kelvin temperatures, where a zero-temperature approximation is
reasonable.

In the next Section we develop the basics of our approach and display
the connection to the well-known ``orthodox'' theory and to the double
junction without charging effects. It follows a discussion of these
results in the third Section.

\section{Time Development}

For the double junction the semi-classical approach results in a
master equation,\cite{ave3,amm2}
\begin{eqnarray}
\label{me}
\frac{{\rm d}\sigma(m,t)}{{\rm d}t} & = & r_1(m-1)\sigma(m-1,t)\\
& & + r_2(m+1)\sigma(m+1,t)\nonumber\\
& & -\big[r_1(m)+r_2(m)\big]\sigma(m,t),\nonumber
\end{eqnarray}
which describes the time dependence of the probability $\sigma(m)$ of
$m$ extra charges on the island. $r_{1,2}$ describe the rates of a
charge tunneling event at the first and second junction. Tunneling
against the bias voltage does not occur for zero temperature. The
number of the charge states $m$ is finite, their energy lies between
the Fermi levels of the banks. We denote the lowest and highest of
these states by ${\rm min}$ and ${\rm max}$, respectively.

With regard to the stochastic process of single-electron tunneling,
the first moment describes the average number of tunneled charges,
$\langle n(t)\rangle$, which is connected with the stationary
electrical current through the double junction, $\langle I\rangle =
e\langle n(t)\rangle/t$, where $e$ is the unit charge. Using the
stationary solution\cite{amm2,seu1} of (\ref{me})
\begin{equation}
\label{ssme}
\sigma(m) = \frac{1}{\cal Z}
\prod_{i={\rm min}}^{m-1}r_1(i)\prod_{i=m+1}^{\rm max}r_2(i),
\end{equation}
with the normalization constant ${\cal Z}$, the stationary current
reads
\begin{equation}
\label{scme}
\langle I\rangle = e\sum_{m={\rm min}}^{\rm max} r_1(m)\sigma(m)
= e\sum_{m={\rm min}}^{\rm max} r_2(m)\sigma(m).\nonumber
\end{equation}
We note that continuity levels the current across the first and the
second junction.

The second moment of the stochastic process under consideration is
connected with the current noise in a double junction. It has been
investigated theoretically in recent years\cite{kor2,han2,mul7} and
its low-frequency limit was measured in experiment.\cite{bir1}

Further moments were - to our knowledge - not subject of theoretical
efforts, however, an approach to the stochastic process of tunneling
through a double junction (without charging effects and at zero
temperature) was performed recently.\cite{jon1} It is based on the 
calculation of the probability $P_n(t)$ of $n$ transmitted charges in
terms of the equations
\begin{eqnarray}
\label{dje}
\frac{{\rm d}P_0(t)}{{\rm d}t} & = & -\gamma_2 P_0(t)\\
\frac{{\rm d}P_n(t)}{{\rm d}t} & = & \gamma_2P_{n-1}(t)-\gamma_2P_n(t).
\nonumber
\end{eqnarray}
$\gamma_{1,2}$ are the transmission probabilities of the two
junctions. According to the Pauli exclusion principle there are only
two different island states that have to be considered, unoccupied
($0$) and occupied ($1$) by one charge. The initial occupation
probability of them is given by\cite{jon1}
\[
P_0^{(0,1)}(t=0) = \frac{\gamma_{2,1}}{\gamma_1+\gamma_2}
\]
and the probability of $n$ transmitted charges at arbitrary time $t$
is\cite{jon1}
\[
P_n(t) = \frac{\gamma_2}{\gamma_1+\gamma_2}P_n^{(0)}(t)
+\frac{\gamma_1}{\gamma_1+\gamma_2}P_n^{(1)}(t).
\]
This problem is solved analytically in Ref.~\onlinecite{jon1}, i.e.\
the transmission probabilities $P_n(t)$ can be calculated. The
description of the stochastic process, however, is much clearer in
terms of its characteristic function $\chi(\lambda,t)$\cite{kam1}
\begin{equation}
\label{cf}
\chi(\lambda,t) = \sum_{n=0}^{\infty} P_n(t)\,{\rm e}^{i\,n\,\lambda},
\end{equation}
from which in turn all moments $\mu_k(t)=\langle n(t)^k\rangle$ of the
stochastic process are obtained
\[
\mu_k(t) = \lim_{\lambda\to0}\big(\frac{\partial}{i\,\partial\lambda}
\big)^k\chi(\lambda,k).
\]
Equivalently, the cumulants follow from the logarithmic expansion of
the characteristic function $\chi(\lambda,t)$.\cite{kam1}

In our approach we include charging effects, however, still assume
zero temperature. Therefore, a finite number of island charge states
$m$ is considered and the description is done in terms of the
probability $\rho(m,n,t)$ of $n$ transmitted charges through both
junctions and simultaneously $m$ excess charges on the island between
them. Thus, the following sum rules hold
\begin{eqnarray}
\label{normrho}
1 & = & \sum_{m={\rm min}}^{\rm max}\sum_{n=0}^{\infty}\rho(m,n,t),\\
\label{sigmarho}
\sigma(m,t) & = & \sum_{n=0}^{\infty}\rho(m,n,t),\\
\label{pnrho}
P_n(t) & = & \sum_{m={\rm min}}^{\rm max}\rho(m,n,t).
\end{eqnarray}

For the time development of $\rho(m,n,t)$ the following first-order
differential equation is found
\begin{eqnarray}
\label{ne}
\frac{{\rm d}\rho(m,n,t)}{{\rm d}t} & = & r_1(m-1)\rho(m-1,n,t)\\
& & + r_2(m+1)\rho(m+1,n-1,t)\nonumber\\
& & -\big[r_1(m)+r_2(m)\big]\rho(m,n,t).\nonumber
\end{eqnarray}
The discussion of this equation and its solution is the main point of
this paper.

The initial condition for the solution of (\ref{ne}) follows from
(\ref{sigmarho}) using $P_0(t=0)=1$,
\begin{eqnarray}
\label{initne}
\rho(m,0,t=0) & = & \sigma(m)\\
\rho(m,n,t=0) & = & 0\mbox{\hspace*{2cm}}n>0.\nonumber
\end{eqnarray}

Summing (\ref{ne}) over $n$ and $m$, respectively, results by the use
of (\ref{sigmarho}) and (\ref{pnrho}) in the former Eqs.~\ref{me}
and~\ref{dje}. Even (\ref{dje}) includes Coulomb effects in our case
indicated by $m$ entering into (\ref{pnrho}). 

Let us consider (\ref{ne}) in the long-time limit ($t\to\infty$)
shortly. Single-electron tunneling is usually considered in this
limit and several simplifications apply. The probability $\rho(m,n,t)$
factorizes $\rho(m,n,t) = P_n(t)\sigma(m)$, where $\sigma(m)$ is the
stationary solution (\ref{ssme}) of the master equation
(\ref{me}). Furthermore, the transmission probabilities become
periodic, $P_{n\pm1}(t) = P_n(t\mp e/\langle I\rangle)$. Under this
conditions the stationary-current formula (\ref{scme}) results from
(\ref{ne}), even for non-zero temperatures.

In general, the assumptions of the last paragraph do not hold and the
solution takes a more complicated form, however, it can be expressed
semi-analytically in terms of a recursion formula. The general
solution of (\ref{ne}) can be written as
\begin{eqnarray}
\label{sne}
\rho(m,n,t) & = & \sum_{k={\rm min}}^m A_{nm}^{nk}\frac{t^n}{n!}
\exp(-\gamma_k t)\\
& & +\sum_{l=0}^{n-1}\sum_{k={\rm min}}^{\rm max}
A_{nm}^{lk}\frac{t^l}{l!}\exp(-\gamma_k t),\nonumber
\end{eqnarray}
where $\gamma_k = r_1(k)+r_2(k)$ is used for abbreviation. Thus, the
solution turns out to be a superposition of different poisson-like
probabilities. The stochastic process as a whole, however, is not
poisson-like. For $n=0$ the second term in (\ref{sne}) vanishes. 

The coefficients $A_{nm}^{lk}$ have to fulfill a recursion relation
which can be derived from (\ref{ne}) by introducing (\ref{sne}). For a
short-hand notation of this recursion relation we make use of the
following decomposition
\[
\frac{1}{(s+\gamma_m)(s+\gamma_k)^{l+1}} = 
\frac{\alpha_{l+1m}^{lk}}{s+\gamma_m}+
\sum_{i=0}^l\frac{\alpha_{im}^{lk}}{(s+\gamma_k)^{i+1}},
\]
which defines the numbers $\alpha_{im}^{lk}$ in a unique way as
solution of a linear equation system. In terms of this notation we 
find

\ifpreprintsty\relax\else\end{multicols}\widetext\rule{\colwidth}{0.4pt}
\hfill\fi

\begin{eqnarray}
\label{anmlk}
A_{nm}^{nk} & = & r_1(m-1)A_{nm-1}^{nk}\alpha_{nm}^{nk}
\mbox{\hspace*{40mm}}k<m\\
A_{nm}^{nm} & = & r_1(m-1)A_{nm-1}^{n-1m}+r_2(m+1)A_{n-1m+1}^{n-1m}
\nonumber\\
A_{nm}^{0m} & = & r_1(m-1)\sum_{k={\rm min}}^{m-1}A_{nm-1}^{nk}
\alpha_{n+1m}^{nk}+r_1(m-1)\sum_{l=0}^{n-1}\sum_{k={\rm min}\atop
k\neq m}^{\rm max}A_{nm-1}^{lk}\alpha_{l+1m}^{lk}\nonumber\\
& & +r_2(m+1)\sum_{k={\rm min}\atop k\neq m}^{m+1}A_{n-1m+1}^{n-1k}
\alpha_{nm}^{n-1k}+r_2(m+1)\sum_{l=0}^{n-2}\sum_{k={\rm min}\atop
k\neq m}^{\rm max}A_{n-1m+1}^{lk}\alpha_{l+1m}^{lk}\nonumber\\
A_{nm}^{lm} & = & r_1(m-1)A_{nm-1}^{l-1m}+r_2(m+1)A_{n-1m+1}^{l-1m}
\mbox{\hspace*{40mm}}1\le l\le n-1\nonumber\\
A_{nm}^{lk} & = & r_1(m-1)A_{nm-1}^{nk}\alpha_{lm}^{nk}
{\rm If}\big[m>k\big]
+r_1(m-1)\sum_{i=l}^{n-1}A_{nm-1}^{ik}\alpha_{lm}^{ik}\nonumber\\
& & +r_2(m+1)A_{n-1m+1}^{n-1k}\alpha_{lm}^{n-1k}
{\rm If}\big[{m+2>k\atop m\neq k}\big]
+r_2(m+1)\sum_{i=l}^{n-2}A_{n-1m+1}^{ik}\alpha_{lm}^{ik}.\nonumber
\end{eqnarray}

\ifpreprintsty\relax\else\hfill
\rule{\colwidth}{0.4pt}\begin{multicols}{2}\narrowtext\fi

For simplicity we made use of the ${\rm If}[\ldots]$ notation: the
corresponding terms contribute only if the given condition is
fulfilled. Eq.~\ref{anmlk} is, in connection with (\ref{sne}), our
main result.

In addition to (\ref{anmlk}), the initialization of $A_{nm}^{lk}$
follows from (\ref{ne}) for $n=0$ using the given initial condition
(\ref{initne}), 
\begin{eqnarray}
\label{a0m0k}
A_{0m}^{0k} & = & -\frac{r_1(m-1)A_{0m-1}^{0k}}{\gamma_k-\gamma_m}
\mbox{\hspace*{20mm}}k<m\\
A_{0m}^{0m} & = & \sigma(m)+r_1(m-1)\sum_{k={\rm min}}^{m-1}
\frac{A_{0m-1}^{0k}}{\gamma_k-\gamma_m}.\nonumber
\end{eqnarray}
Eq.~\ref{a0m0k} does not hold in the very symmetric case. Then, the
solution might derived from the given set of equation in terms of a
limit procedure $\gamma_k\to\gamma_m$.  

Eqs.~\ref{anmlk} and~\ref{a0m0k} allow the construction of all
$A_{nm}^{lk}$. These coefficients tell the important contributions
within (\ref{sne}) from the unimportant ones. If a sole term 
dominates the stochastic process will approach a Poisson process,
similar to the case without Coulomb effects.\cite{jon1} In general,
however, more terms contribute and the total process is a
superposition of Poisson processes. The result (\ref{anmlk}), however,
does not seem to provide clear guidance to a simplification by the
selection of relevant terms. Even if the given recursion relation
(\ref{anmlk}) looks clumsy its numeric implementation is neat.

\section{Discussion}

Owing to the complicated expressions (\ref{anmlk}) we discuss our
results by means of numerical computations. The most important
question concerns the time domain where (\ref{sne}) is valid. Due to
the derivation of the rates $r_{1,2}(m)$ by Fermi's Golden Rule, the
time scale is limited to $t\gg R_{1,2}C_{1,2}$. On the other hand, for
very long times the system will approach the mentioned stationary
state of the island and the investigation looses its thrill. 

\ifpreprintsty\relax\else
\ifx\epsfxsize\undefined\relax\else
\begin{figure}
\epsfxsize=\columnwidth
\epsffile{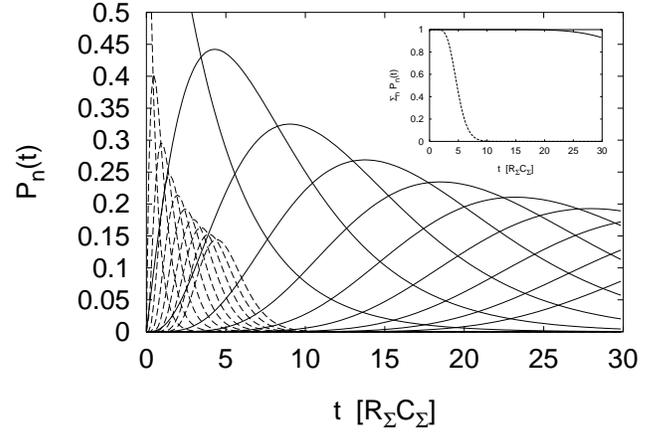}
\caption{Transmission probability of charge through a double junction
with (solid lines) and without (dashed lines) Coulomb effects for the
first ten transmitted charges. The inset shows displays the sum of
these ten probabilities.}
\label{fig1}
\end{figure}
\fi
\fi

In Fig.~\ref{fig1} we show the results for $P_n(t)$ for
$n=0\ldots9$. The parameters of the double junction under
investigation are $R_1=R_2=100{\rm k}\Omega$, $C_1=10{\rm aF}$,
$C_2=101{\rm aF}$. The solid lines show the data corresponding to
(\ref{sne}), whereas the dashed lines reflect the results without
consideration of the charging effects according to
Ref.~\onlinecite{jon1}. 

Since not shown in the figure let us mention the initial condition
$P_0(t=0)=1$, which is fulfilled by both sets of transmission
probabilities. The suppression of the charge transmission by Coulomb
interaction, however, is clearly observed in Fig.~\ref{fig1}: The
probability $P_n(t)$ with charging effects extends over a wider time
range than the same probability without charging for a given number
$n$ of transmitted charges. Thus, for a fixed time $t$ less charges
are transmitted in former case and the conduction is suppressed.

In the inset the probability
\[
\sum_{n=0}^9 P_n(t)
\]
is shown. For short times it approximates the total probability $1$,
whereas later neglected states get more important. We note that
the first ten $P_n(t)$ are sufficient to describe times up to
$\approx20\,R_{\Sigma}C_{\Sigma}$ in the case with Coulomb
interaction, but only up to $\approx2\,R_{\Sigma}C_{\Sigma}$ in the
case without charging. For short times the applied description of the
system fails, however. 

\ifpreprintsty\relax\else
\ifx\epsfxsize\undefined\relax\else
\begin{figure}
\epsfxsize=\columnwidth
\epsffile{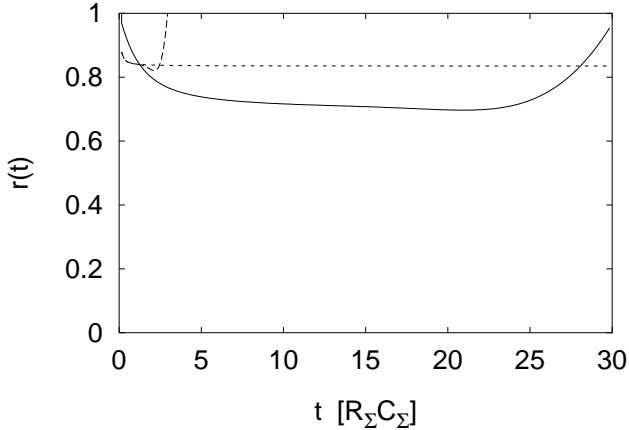}
\caption{Fano factor $r(t)$ for a double junction with and without
Coulomb interaction (solid and dashed line, respectively). The
short-dashed line corresponds to the long-time limit of the case
without charging effects.}
\label{fig2}
\end{figure}
\fi
\fi

In order to apply the solution (\ref{sne}), we plot the Fano factor
\[
r(t) = \frac{\mu_2(t)}{\mu_1(t)}-\mu_1(t)
\]
for the case with and without Coulomb interaction in
Fig.~\ref{fig2}. The Fano factor describes the correlation of
single tunneling events. Small $r(t)$ indicate a large degree of
correlation. Therefore, a high Fano factor is expected at $t=0$,
whereas lateron it should approach a lower stationary
limit.\cite{kor2,han2}

Again, our computation makes use of the first ten $P_n(t)$ only. This
corresponds to Fig.~\ref{fig1}. Owing to the limited number $n$ the
plotted approximation fails at longer times, i.e.\ when the total
probability (as shown in the inset of Fig.~\ref{fig1}) differs from
unity. Besides this flaw the expected behavior of the Fano factor is
recovered. For the case without Coulomb interaction we plot the
long-time limit of Ref.~\onlinecite{jon1} as well.

Coulomb interaction favors correlated tunneling. This is seen by the
lower value of $r(t)$ in comparison to the stationary case without
Coulomb interaction. The degree of correlation depends on the symmetry
of the double junction [\ldots]. It gets maximal in case of a
symmetric double junction ($R_1=R_2$, $C_1=C_2$), where the Fano
factor reaches $1/2$. The double junction under consideration,
however, is rather asymmetric. Thus, the decrease of the Fano factor
is small.

For the studied case with Coulomb interaction noticeable changes in
$r(t)$ are observed even in the time range of
$t\approx10R_{\Sigma}C_{\Sigma}$. This is well above the critical
short-time range, where the approach in terms of the Golden Rule is
insufficient. This fact supports justification of the presented
approach. 

\ifpreprintsty\relax\else
\ifx\epsfxsize\undefined\relax\else
\begin{figure}
\epsfxsize=\columnwidth
\epsffile{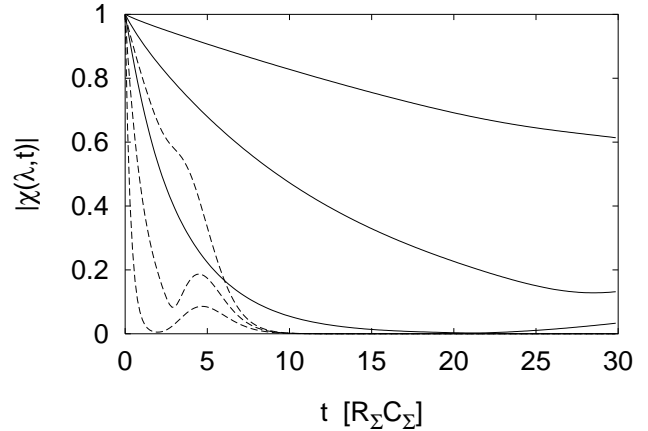}
\caption{Absolute value of the characteristic function
$|\chi(\lambda,t)|$ for three different values of $\lambda$ ($0.5$,
$1.0$, $2.0$, from top to bottom). The double junction parameters
correspond to the other figures. Again, solid and dashed lines
indicate the computation with and without Coulomb energy, respectively.}
\label{fig3}
\end{figure}
\fi
\fi

Finally, in Fig.~\ref{fig3} we display the time-dependence of the
characteristic function $\chi(\lambda,t)$ of Eq.~\ref{cf}. As
mentioned above, it ``characterizes'' the stochastic process of
electron tunneling through the double junction. Again, we use the
probabilities of the first ten transmitted charges for
computation. Therefore, the relevant time range is here similar to
that of Fig.~\ref{fig1} and Fig.~\ref{fig2}. This plot shows that the
presented recursion relation scheme can be used to compute further
properties of the underlying stochastic process.

\section{Conclusion}

In this paper we have studied the stochastic process of
single-electron tunneling in an ultrasmall normal-metal double-tunnel
junction at zero temperature. We have introduced an equation,
(\ref{ne}), which enables the investigation of transmission
probabilities of charge corresponding to the ``orthodox theory'' of
single-electron tunneling. The time-dependent solution of this equation
is derived in terms of a recursion relation, (\ref{anmlk}), with the
appropriate initialization (\ref{a0m0k}). The results are discussed in
comparison to the known investigation of the double junction without
charging effects.\cite{jon1} We observe a suppression of tunneling due
to Coulomb repulsion and an enhanced correlation. The recursion
relation provides a suitable tool for computation of stochastic
properties of the double junction. The absolute value of the
characteristic function $\chi(\lambda,t)$ serves as an example.

\section*{Acknowledgment}

Hospitality at the Friedrich--Schiller--Universit{\"a}t Jena, Germany,
is gratefully acknowledged.


\ifpreprintsty\narrowtext\else\relax\fi

\ifx\epsfxsize\undefined
\begin{figure}
\caption{Transmission probability of charge through a double junction
with (solid lines) and without (dashed lines) Coulomb effects for the
first ten transmitted charges. The inset shows displays the sum of
these ten probabilities.}
\label{fig1}
\end{figure}

\begin{figure}
\caption{Fano factor $r(t)$ for a double junction with and without
Coulomb interaction (solid and dashed line, respectively). The
long-dashed line corresponds to the long-time limit of the case
without charging effects.}
\label{fig2}
\end{figure}

\begin{figure}
\caption{Absolute value of the characteristic function
$|\chi(\lambda,t)|$ for three different values of $\lambda$ ($0.5$,
$1.0$, $2.0$, from top to bottom). The double junction parameters
correspond to the other figures. Again, solid and dashed lines
indicate the computation with and without Coulomb energy, respectively.}
\label{fig3}
\end{figure}
\else\relax\fi

\ifpreprintsty\relax\else\end{multicols}\fi

\end{document}